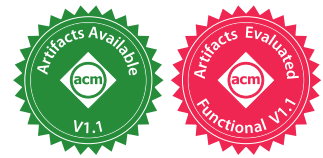

# Perfect Is the Enemy of Test Oracle

Ali Reza Ibrahimzada
University of Illinois Urbana-Champaign, USA
alirezai@illinois.edu

Yigit Varli
Middle East Technical University, Turkey
yigit.varli@metu.edu.tr

Dilara Tekinoglu
University of Massachusetts Amherst, USA
dtekinoglu@umass.edu

Reyhaneh Jabbarvand
University of Illinois Urbana-Champaign, USA
reyhaneh@illinois.edu

## ABSTRACT

Automation of test oracles is one of the most challenging facets of software testing, but remains comparatively less addressed compared to automated test input generation. Test oracles rely on a ground-truth that can distinguish between the correct and buggy behavior to determine whether a test fails (detects a bug) or passes. What makes the oracle problem challenging and undecidable is the assumption that the ground-truth should know the exact expected, correct, or buggy behavior. However, we argue that one can still build an accurate oracle without knowing the exact correct or buggy behavior, but how these two might differ. This paper presents SEER, a learning-based approach that in the absence of test assertions or other types of oracle, can determine whether a unit test passes or fails on a given method under test (MUT). To build the ground-truth, SEER jointly embeds unit tests and the implementation of MUTs into a unified vector space, in such a way that the neural representation of tests are similar to that of MUTs they pass on them, but dissimilar to MUTs they fail on them. The classifier built on top of this vector representation serves as the oracle to generate "fail" labels, when test inputs detect a bug in MUT or "pass" labels, otherwise. Our extensive experiments on applying SEER to more than 5K unit tests from a diverse set of open-source Java projects show that the produced oracle is (1) *effective* in predicting the fail or pass labels, achieving an overall accuracy, precision, recall, and F1 measure of 93%, 86%, 94%, and 90%, (2) *generalizable*, predicting the labels for the unit test of projects that were not in training or validation set with negligible performance drop, and (3) *efficient*, detecting the existence of bugs in only 6.5 milliseconds on average. Moreover, by interpreting the neural model and looking at it beyond a closed-box solution, we confirm that the oracle is valid, i.e., it predicts the labels through learning relevant features.

## 1 INTRODUCTION

A unit test similar to the example in Figure 1 consists of four main components: test input ("e1", "e2", and "e3"), MUT invocation (obj.sort()), test output, and test oracle (assertEquals). Given a ground-truth that knows the program's expected correct or buggy behavior for given inputs, oracles can determine test results, i.e., whether a test passes or fails. For example, the ground-truth in the example of Figure 1 identifies the sorted output for given inputs to be "e1,e2,e3". Consequently, the assertion oracle checks if the produced output matches the expected one to generate the test result. Construction of the ground-truth can be a challenging task. Furthermore, the absence of automated ground-truths demands humans to decide whether the generated outputs are correct, demonstrating a significant bottleneck that inhibits absolute test automation [6].

```
@Test
public void testAdd() {
  ExampleObject obj = new ExampleObject();
  obj.add("e1");
  obj.add("e2");
  obj.add("e3");
  String output = obj.sort();
  Assert.assertEquals(output,"e1,e2,e3");
}
```

**Figure 1: A simple JUnit test consists of four main components: input, MUT invocation, output, and assertion**

To automatically build the ground-truth for test oracles, traditional and machine learning-based techniques rely on existing or derived formal specifications [18], assertions [25, 51, 72, 77, 80], program invariants [23, 88], and metamorphic relations [13, 14, 87, 90] for identifying the correct behavior. Some other techniques determine the patterns corresponding to specific types of bugs observed during test execution as an indicator of the buggy behavior [35, 45]. The commonality between these techniques is their emphasis on identifying the *exact* correct or buggy behavior to build the ground-truth. However, identifying the exact behavior and output is an undecidable problem; thereby, such techniques only partially validate the program. That is, the program can only be validated under a subset of test inputs with known expected outputs, or limited properties of the program determined by the invariants or metamorphic relations can be validated. While partial oracles enhance test automation to some degree, they may not guarantee the existence of a successful test driver.

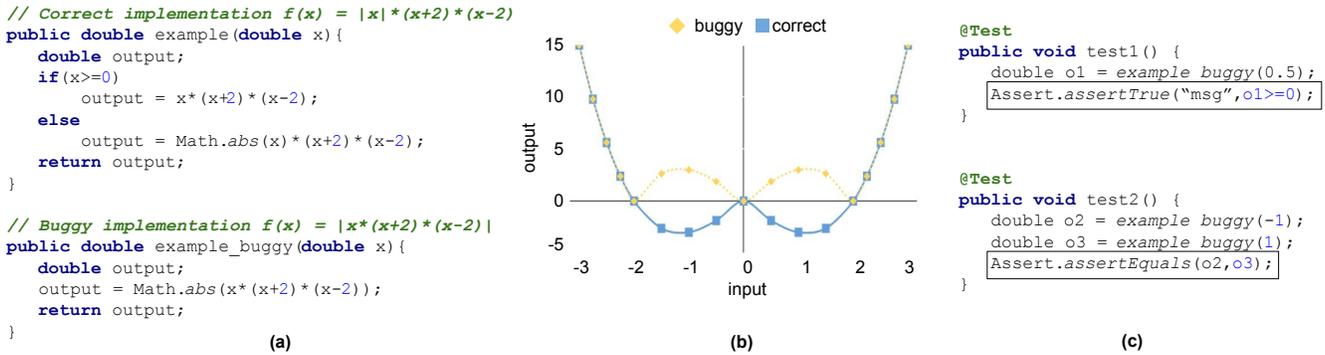

Figure 2: (a) correct (top) and buggy (bottom) implementations of the mathematical function $f(x) = |x| \times (x+2) \times (x-2)$, (b) visualization of the correct and buggy implementations behavior, (c) JUnit tests with assertions to assess correctness of the buggy implementation

The key insight in our research is that one can still create an accurate test oracle without knowing the *explicit relationship* between inputs and outputs under correct or buggy behavior. Instead, the ground-truth will determine how *different* the test inputs are *correlated* to outputs under the correct and buggy behavior. An oracle based on this ground-truth eliminates the need for assert statements and identifying the exact expected output in assertions, enhancing unit testing to a great extent [77].

In this paper, we present SEER [1], an automated oracle to predict unit test results. Specifically, given a pair of <$t_i$, $m_i$>, where $t_i$ represents a unit test *without any assertions*, and $m_i$ denotes the implementation of MUT, SEER automatically determines whether the test passes or fails on the MUT. To construct the ground-truth, SEER leverages joint embedding to distinguish between the neural representation of correct and buggy MUTs. A classifier on top of this embedding learns the correlation between inputs and outputs to predict test results. This paper makes the following contributions:

- A novel domain-specific joint embedding of the unit tests and MUTs, which semantically separates MUTs' neural representations based on whether unit tests pass or fail on them.
- Design of an *interpretable* DL model that serves as a test oracle to generate passing or failing labels for unit tests without assertions. The interpretability enables us to go beyond the usage of DL as a closed-box technique and verify if the model predicts labels by looking at the relevant tokens in the implementation of MUTs. While it is out of the scope of this paper, the relevant tokens involved in the model's decision can be further used by developers to localize the detected bugs.
- An extensive empirical evaluation on widely used open-source Java programs demonstrating that SEER is (1) *effective*—achieves an overall accuracy, precision, recall, and F1 measure of 93%, 86%, 94%, and 90%, (2) *generalizable*—predicting the labels for the unit test of projects that were not in training or validation set with negligible performance drop, and (3) *efficient*—once trained, it detects the existence of bugs in only 6.5 milliseconds on average. SEER's implementation and artifacts are publicly available [2].

The remainder of this paper is organized as follows. Section 2 illustrates a motivating example. Section 3 provides an overview of SEER, while Section 4 describes details of the proposed technique. Section 5 presents the evaluation results. The paper concludes with a discussion of the related research and future work.

## 2 ILLUSTRATIVE EXAMPLE

To illustrate the limitations of prior work and explain the intuition behind our research, we use two code examples shown in Figure 2-a. The code snippet on the top is the *correct* implementation of a mathematical function that computes the output as $|x| \times (x+2) \times (x-2)$. The buggy version, on the other hand, computes the output as $|x \times (x+2) \times (x-2)|$ due to the displacement of a single parenthesis to the end of the assignment instead of after variable x.

Formally speaking, the behavior of a code is a function $B : I \rightarrow O$ that maps inputs in $I$ to corresponding outputs in $O$. In our example, the mapping functions representing the explicit behavior of correct and buggy implementations are depicted as blue and yellow graphs in Figure 2-b. If such a function is known, an oracle can use it as a ground-truth to distinguish the buggy and correct behavior. However, in reality, MUTs take multiple complex inputs, e.g., arrays and user-defined objects, resulting in n-dimensional mappings between inputs and outputs that are infeasible to determine. Therefore, test oracles rely on partial ground-truths. No matter how we build the ground-truth, the oracle's decision for test inputs belonging to $(-2, 0) \cup (0, 2)$ should be *"fail"* due to different behavior of the correct and buggy implementations in this range.

Suppose that we have two JUnit tests shown in Figure 2-c to assess the correctness of the example_buggy method. The ground-truth for identifying the expected output in the assert statement of test1 is based on dynamic invariant detection, while the ground-truth for the assert statement in test2 is based on a metamorphic relation $f(x) = f(-x)$. Dynamic invariant detection techniques rely on the execution traces of the *existing code*. Since our MUT is buggy, the invariant only captures properties of the buggy behavior, i.e., $output \geq 0$. By checking the generated invariant in the assertion,

---
[1]A person who can see what the future holds through supernatural insight.

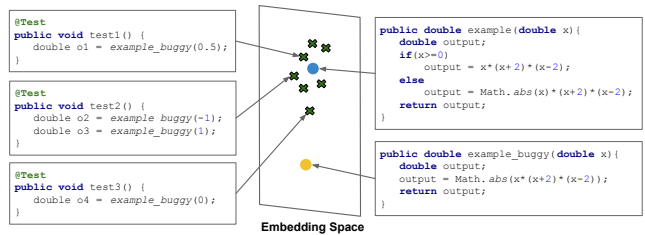

**Figure 3: The intuition behind the joint embedding of tests and MUTs with the goal of separating the representation of buggy and correct MUTs with respect to tests**

the test *passes* for $x = 0.5 \in (-2, 0) \cup (0, 2)$, while it should *fail* to demonstrate the bug. Similarly, the metamorphic relation of $f(x) = f(-x)$ holds for both correct and buggy implementations; thereby, the assertion wrongly decides the test inputs in the non-overlapping range as passed.

This example shows that identifying the explicit correct (or buggy) behavior to build a ground-truth, which has been the focus of prior work, has notable limitations. In this research, instead of realizing how the inputs are *explicitly related* to outputs under either correct or buggy behavior, we aim to learn how inputs are *differently correlated* to outputs for failing and passing pairs of <$t_i, m_i$>. Here, $t_i$ represents a unit test without any assertions, and $m_i$ denotes the implementation of the MUT. To that end, SEER learns the vector representation of both MUT and test, so that the tests have a similar vector representation to the MUTs they pass on them, but dissimilar vector representation to MUTs they can reveal their bug, i.e., fail on them. Such joint embedding separates the representation of buggy and correct MUTs in the n-dimensional vector space.

Figure 3 shows the intuition behind the joint embedding in SEER. Here, since test inputs in test1 and test2 can reveal the bug in example_buggy, they are closer to the correct MUT and farther from the buggy MUT in the embedding space, i.e., they have a similar vector representation as correct MUT but dissimilar from buggy MUT. On the other hand, test3 that cannot reveal the bug and should pass on both the buggy and correct MUTs, has the same distance from the correct and buggy MUTs in the embedding space. Compared to Figure 2-b, there is no explicit relationship between test inputs and outputs under correct or buggy behavior, but the embedding representation of correct and buggy MUTs are distinguished based on how the output they generate are correlated differently to the inputs of passing or failing tests.

## 3 FRAMEWORK OVERVIEW

Figure 4 provides an overview of SEER framework consisting of four major components: (1) *Method Extractor*, (2) *Dataset Augmentor*, (3) *Learning Module*, and (4) *Interpreter*. SEER requires a high-quality and large dataset of <$t_i, m_i$> instances to train the oracle. Given a set of programs and their corresponding test suites, the *Method Extractor* component builds such a dataset by extracting the implementation of MUT invocations in tests through a lightweight static analysis (details in §4.1). At the next step, *Method Extractor* creates labeled tuples in the form of ⟨$t_i, m_i, l_i$⟩, where $t_i$ represents a unit test, $m_i$ denotes the implementation of MUT, and $l_i$ shows the test result outcome, which could be $P$ (pass) or $F$ (fail). *Dataset Augmentor* component then takes the generated dataset and augments it with additional instances to diversify the bugs and account for imbalanced labels (details in §4.3).

Once the training dataset is ready, SEER feeds it to the *Learning Module* to train the oracle through two phases. In the Phase 1 training, the *Learning Module* learns the vector representation of the test $t_i$ and the MUT $m_i$ through joint embedding by minimizing the distance among passing tuples ⟨$t_i, m_i, P$⟩, while maximizing the distance among failing tuples ⟨$t_i, m_i, F$⟩ (detail in §4.2). As a result, the vector representation of a test is similar to the MUTs it passes on them, but dissimilar to the MUTs it fails on. After learning the discriminative vector representations, *Learning Module* leverages transfer learning [52] and trains a classifier on top of the embedding network, which serves as our test oracle. To predict the label, SEER takes a unit test and the program under test as an input and extracts the implementation of invoked MUT(s). Given the produced pair of ⟨$t_i, m_i$⟩, the embedding network first computes their vector representations, and the classifier predicts the label, indicating whether a test passes on the given MUT or fails.

SEER goes beyond the use of DL as a closed-box approach and interprets the learned model for two purposes: (1) to verify if the embedding does its job in separating the representation of MUTs based on whether a test passes or fails on them, and (2) to verify the validity of the model by checking if the code tokens that impacted the oracle's decision are relevant (§4.4). In the next section, we describe the details of SEER's components.

## 4 SEER

This section will first explain how to prepare the inputs to the *Learning Module*, followed by the details about SEER's neural architecture, dataset curation, and model interpretation.

### 4.1 Method Extractor

SEER's *Learning Module* requires labeled pairs of <$t_i, m_i$> to realize the correlation between the inputs (provided by unit tests $t_i$) and outputs (produced by MUT $m_i$). Given a test suite $T = \{t_1, t_2, \ldots, t_n\}$ consists of $n$ unit tests and the program under test $P = \{m_1, m_2, \ldots, m_k\}$ consists of $k$ developer-written methods[2], *Method Extractor* extracts $m_i$, the implementation of a MUT directly called in the body of the unit test $t_i$.

Method extraction can be performed statically, i.e., extracting the whole body of the MUTs regardless of the statements covered by a given test, or dynamically, i.e., only considering the executed statements by a test. While the latter is more intuitive in helping the model focus on the executed lines for predicting test verdicts, recent studies have shown that neural models learn the semantics of the code and context more effectively if provided global information [34, 76]. Consequently, the *Model Extractor* performs a lightweight flow-sensitive analysis on a given unit test $t_i$, identifies the MethodInvocation that belongs to the program under test, and extracts the corresponding method signature and the body. If a test invokes multiple methods, *Model Extractor* concatenates the extracted information for the MUTs in the order of invocation. In the illustrative example of Figure 2, *Method Extractor* identifies

---
[2]We exclude third-party APIs as their code may not be available.

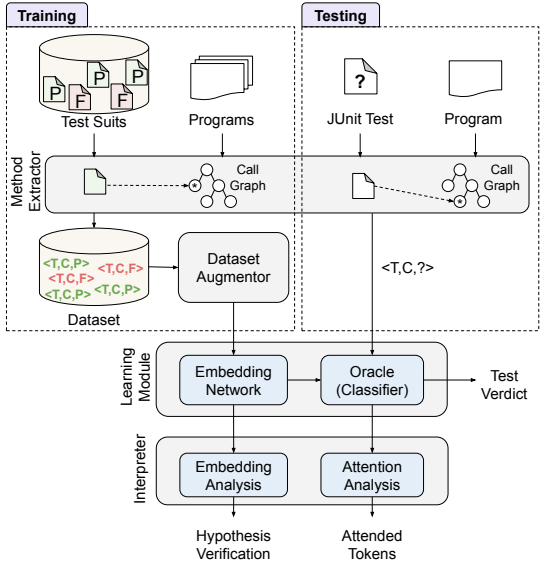

Figure 4: Overview of the SEER framework

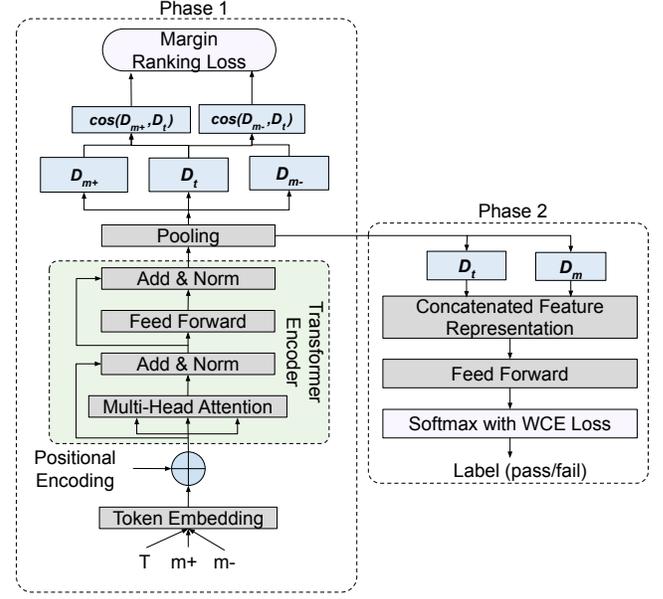

Figure 5: Overall architecture of SEER

example_buggy method as $m_i$ and extracts the whole text of the method, including the method signature and body.

## 4.2 Learning Module

The neural architecture of SEER is shown in Figure 5. Learning the neural model that serves as an oracle happens in two phases. In Phase 1 training, SEER learns the vector representation of unit tests and MUTs, in such a way that the representation of buggy and correct MUTs are different. At the next step, the representation of $<t_i, m_i>$ pairs will be fed into a classifier, helping SEER to learn the correlation between inputs provided by $t_i$ and outputs produced by $m_i$. Since the produced representation of buggy and correct MUTs are different, the oracle ultimately learns how differently the inputs are correlated to outputs under the correct and buggy behavior.

During the Phase 1 training, SEER learns the vector representation of unit tests and MUTs through joint embedding [84]. Joint embedding, also known as multi-modal embedding, has been widely used to embed heterogeneous data into a unified vector space so that semantically similar concepts across the two modalities reside closer in the embedding space. For example, in computer vision, researchers have used Convolutional Neural Network (CNN) and Recurrent Neural Network (RNN) to jointly embed images and text into the same vector space for labeling images [39].

We adopt the concept of joint-embedding in our problem to semantically separate the representation of correct and buggy MUTs concerning the result of tests. Specifically, we hypothesize that by jointly embedding the unit tests and MUTs into a unified vector space, so that tests have similar vector representations to the MUTs they pass on them but are different from the MUTs they fail on them, the resulting embedding separates the representation of correct and buggy MUTs. The joint embedding of unit test, $t_i$, and implementation of MUT, $m_i$, can be formulated as follows:

$$m_i \xrightarrow{\phi} D_{m_i} \rightarrow J(D_{m_i}, D_{t_i}) \leftarrow D_{t_i} \xleftarrow{\psi} t_i$$

where $\phi$ is an embedding function to map $m_i$ into a d-dimensional vector space $D$; $\psi$ is an embedding function to map $t_i$ into the same vector space $D$; and $J(,)$ is a similarity measure to score the matching degrees of $D_{m_i}$ and $D_{t_i}$ in order to put $m_i$ and $t_i$ closer or farther in the embedding space. SEER uses the cosine similarity metric to measure the similarity between the vector representations of $t_i$ and $m_i$. A small cosine similarity means two vectors are closer together in d-dimensional embedding spaces, while a bigger cosine similarity means the vectors point to different angles, i.e., are farther from each other in the embedding space. To learn and semantically separate the representation of correct and buggy $m_i$s with respects to the test results, SEER minimizes the ranking loss as follows:

$$\mathcal{L}(\theta) = \sum_{\langle t_i, m_i+, m_i-\rangle} max(cos(D_{t_i} - D_{m_i+}) - cos(D_{t_i} - D_{m_i-}) + \alpha, 0)$$

where $D_{t_i}$ is the vector representation of $t_i$, $D_{m_i+}$ is the vector representation of a MUT that $t_i$ passes on it, and $D_{m_i-}$ is the vector representation of a MUT that $t_i$ fails on it. The $\theta$ and $\alpha$ represent model parameters and a constant margin value, respectively. Intuitively, by minimizing the margin ranking loss, SEER learns to minimize the distance between $D_{t_i}$ and $D_{m_i+}$, while maximizing the distance between $D_{t_i}$ and $D_{m_i-}$.

After learning the vector representation of tests, $D_{t_i}$, and MUTs, $D_{m_i}$, in Phase 1, SEER concatenates them to create a single continuous feature representation for the $<t_i, m_i>$ pair. The resulting combined feature vector is fed into a series of fully connected layers in Phase 2 training to decode the learned features into a specific target class, i.e., pass or fail.

## 4.3 Dataset Curation

Training of SEER requires a large and high-quality dataset, i.e., a dataset consists of passing and failing $\langle t_i, c_i \rangle$ pairs representing a diverse set of bugs across different projects. To construct the dataset, we started with the Defects4J [1], which is a collection of reproducible bugs in large and widely-used Java projects. Each bug in this dataset is accompanied by the buggy and fixed versions of subject programs as well as developer-written passing and failing tests. Our rationales to build the dataset based on Defects4J are: (1) the bugs are isolated and reproducible, making it easier for the neural model to learn relevant features; and (2) it contains failing developer-written tests, helping with the generation of a balanced dataset, since the automated generation of failure-triggering tests using Randoop [51] and EvoSuite [25] is not guaranteed.

A significant limitation of Defects4J dataset is the complexity of the bugs, i.e., the majority of bugs involve only one statement in the code. This issue can degrade the performance of SEER in two ways. First of all, the model may treat small changes in the code as noise and may not include them in learning, achieving a lower performance [33, 79]. More importantly, a model trained on simple bugs may not generalize to complex, more realistic bugs. Inspired by the power of mutation testing in curating high-quality training datasets [35] and the fact that higher-order mutants [37] are more representative of complex bugs, *Dataset Augmentor* component of SEER takes a passing $\langle t_i, m_i \rangle$ pair as input and mutates the MUT repeatedly at different locations to generate *higher-order* mutants.

Algorithm 1 explains our dataset augmentation process. The algorithm takes the *MUT* and *order*—the maximum number of times we mutate a given MUT—as an input and generates a higher-order mutant, *HOM*. To that end, it first identifies unique pairs of *Mutables* = $\langle op, loc \rangle$ that demonstrate the locations *loc* in MUT where a mutation operator *op* can be applied (Line 1). Next, it mutates the MUT once at a time using these pairs (Lines 4-5). After each mutation, the algorithm checks whether or not the generated mutant is compilable (Line 7). The algorithm continues the mutation using the next *mutable* (Lines 8-9) if the mutant is compilable. Otherwise, it reverts the mutation and terminates with the produced *HOM* (Lines 10-11). Mutation continues until the MUT is mutated *order* times or it has been mutated at all the mutable locations.

### 4.4 Interpretation

Without interpretation, one cannot trust the performance of ML, and specifically DL models, as their learning depends on millions of parameters. Specifically, such intelligent models can create unrealistically good predictions, but based on learning from irrelevant features due to the noise in the dataset or data leakage problem [40]. To ensure the trustworthiness of SEER, we validate the following two hypotheses by interpreting the *Learning Module*: *Hypothesis 1*. The oracle looks at relevant tokens in the MUT to predict a test result; and (2) *Hypothesis 2*. The embedding network separates buggy and correct MUTs in the embedding space by distinguishing their vector representations. We will discuss the details of *Attention Analysis* and *Embedding Analysis* to investigate the correctness of *Hypothesis 1* and *Hypothesis 2*, respectively.

#### 4.4.1 Attention Analysis.
Attention mechanism [15], which was initially proposed to overcome the *long sequence problem* in Recurrent Neural Networks (RNNs), is a method for helping DL models to identify the importance of single features in a feature sequence as they perform their tasks. Attention mechanism serves two purposes in neural architectures; first, it helps with the model's performance.

**Algorithm 1:** Dataset Augmentation Algorithm
**Input:** MUT, order
**Output:** A Higher Order Mutant of MUT (HOM)
1 *Mutables* ← getUniqueMutables (*MUT*)
2 *counter* ← *order*
3 *HOM* ← *MUT*
4 **foreach** *mutable* ∈ *Mutables* **do**
5    *mutant* ← mutate (*MUT*, *mutable.op*, *mutable.loc*)
6    **if** *counter* > 0 **then**
7       **if** *isCompilable(mutant)* **then**
8          *HOM* ← *mutant*
9          *counter* ← *counter* − 1
10       **else**
11          **return** *HOM*
12    **else**
13       *break*
14 **return** *HOM*

More importantly, it has been extensively used to resolve the interpretability of deep neural models. The initial implementations of the Attention mechanism were neural layers between the encoder and decoder components in the neural architecture, producing attention weight vectors, $\overrightarrow{AT} = \{w_0, \ldots, w_n\}$, as an output. In the context of neural code analysis, $w_i$ is the probability that given a statement with $n$ tokens, how important is the token at location $i$ in the statement when predicting a label. The higher the Attention weight for a feature, the more the model attends to it when making a prediction.

SEER uses Multi-Head Attention [15, 74], also known as Self Attention, to consider the relative importance of a token in the statement when learning. The output of a Self Attention layer is an $n \times n$ matrix $SA = [[w_{0_0}, \ldots, w_{0_n}], \ldots, [w_{n_0}, \ldots, w_{n_n}]]$, where $w_{ij}$ represents the how important is the token at location $i$ given a specific token at location $j$. Figure 6 shows the difference between these two Attention mechanisms for the buggy statement in our

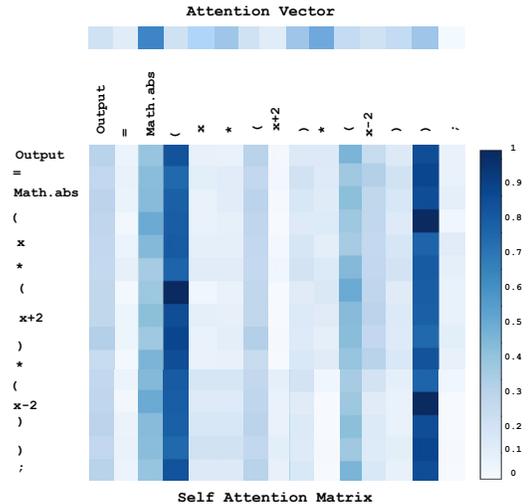

**Figure 6: The importance of Self Attention in the oracle problem**

**Algorithm 2:** Attention Analysis

**Input:** MUT's tokens $\overrightarrow{Tkn} = \{c_0, \ldots, c_n\}$
**Input:** MUT's statements $\overrightarrow{Smt} = \{s_0, \ldots, s_m\}$
**Input:** $SA = [[w_{0_0}, \ldots, w_{0_n}], \ldots, [w_{n_0}, \ldots, w_{n_n}]]$
**Input:** Attention threshold $k$
**Output:** Attended tokens $ATkn$, Attended statements $ASmt$

1  $ATkn \leftarrow \emptyset$
2  $ASmt \leftarrow \emptyset$
3  **foreach** $row = [w_{i_0}, \ldots, w_{i_n}] \in SA$ **do**
       // $localATkn = \{\langle c_i, ind_i \rangle | ind_i$ is index of $c_i$ in $\overrightarrow{Tkn}\}$
4     $localATkn \leftarrow getMostAttended(row, Tkn, k)$
5     **foreach** $\langle c_i, ind_i \rangle \in localATkn$ **do**
6        **if** $\neg ATkn.contains(\langle c_i, ind_i \rangle)$ **then**
7           $ATkn \leftarrow ATkn \cup c_i$
8  **foreach** $s_i \in Smt$ **do**
9     **if** $|s_i \cap ATkn| > k$ **then**
10       $ASmt \leftarrow ASmt \cup s_i$

illustrative example (for the sake of space and readability, Figure 6 shows only a subset of SA corresponding to the buggy statement). (Figure 2-a). As shown in this figure, Self Attention is more successful at capturing the importance of closing parenthesis with respect to open ones compared to the traditional Attention mechanism. Algorithm 2 presents SEER's approach for Attention analysis, i.e., analyzing the *SA* matrix to identify which tokens and statements attended the most in the MUT to predict the test result.

For a given pair of $\langle t_i, m_i \rangle$, Algorithm 2 takes MUT's tokens, $\{c_0, \ldots, c_n\}$, Self Attention matrix, $SA = [[w_{0_0}, \ldots, w_{0_n}], \ldots, [w_{n_0}, \ldots, w_{n_n}]]$, and Attention threshold value, $k$, as an input to identify the set of attended tokens, $ATkn$ and attended statements $ASmt$, as outputs. To that end, the Algorithm traverses $SA$ matrix row by row (Lines 3-7), identifies the top $k\%$ most attended tokens—top $k$ tokens with the highest Attention weight value (Line 4), and merges the attended tokens per each row for the entire matrix along with their corresponding indices (Lines 5-7). The outcome of merge is the set of attended tokens, $ATkn$. When merging, considering the indices is specifically important, as similar tokens at different indices might be attended differently. In the example of Figure 6, while the token "(" appears multiple times in SA at different indices, its highest attention is in the last index.

Finally, the Algorithm iterates over MUT's statements, $\{s_0, \ldots, s_m\}$, and determines the statements that $k\%$ of their tokens overlap with the attended tokens in $ATkn$ (Lines 8-10). Such statements indicate the buggy statements in cases that the predicted label for a $\langle t_i, m_i \rangle$ pair is "fail". The intuition here is that since the number of buggy lines is limited, according to the *Generalized Pigeonhole Principle* [10], there is at least one statement with more than $k\%$ tokens among attended tokens $ATkn$.

*4.4.2 Embedding Analysis.* SEER relies on visualization techniques to validate the separation of the buggy and correct MUT representations in the embedding space. Given the high dimensionality of embedding vectors, however, the first step in the embedding analysis is reducing the dimension of representations. Dimensionality reduction algorithms such as PCA [53], LDA [70], and tSNE [29]

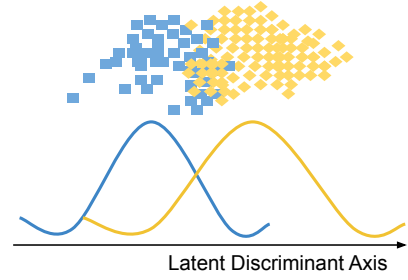

**Figure 7: The intuition behind LDA dimensionality reduction**

concentrate on placing dissimilar data points far apart in a lower dimension representation.

Among the most popular dimensionality reduction algorithms, SEER uses Linear Discriminant Analysis (LDA), as it recognizes the class labels and maximizes the separation between classes during the dimensionality reduction. Figure 7 shows the intuition behind how LDA performs high-dimensionality reduction. The scattered plots represent the distribution of buggy (yellow) and correct (blue) MUTs in high-dimensional embedding space, and the curves represent that of in the lower dimension. As demonstrated by Figure 7, if the distributions of reduced-dimension instances of two classes overlap, they are not separated correctly in the higher dimension. Otherwise, the instances of buggy and correct classes are separated in the high-dimensional embedding space.

## 5 EVALUATION

To evaluate effectiveness of SEER, we investigate the following research questions:
**RQ1:** *Effectiveness.* How effective are the proposed techniques in predicting accurate passing or failing test labels? What type of bugs the proposed oracle can detect, and what bugs are harder for the oracle to detect?
**RQ2:** *Generalization.* To what extent the proposed technique can predict test labels for the Java projects it has not been trained on?
**RQ3:** *Interpretation.* Can embedding truly distinguish between the representation of MUTs for passing and failing <test,MUT> pairs? What features impact the oracle's decision?
**RQ4:** *Performance.* What are the performance characteristics of the proposed technique?

### 5.1 Experimental Setup

We will explain the details of our experimental setup for the sake of reproducibility. Moroever, we have made all artifacts of SEER publicly available on GitHub [2].

*5.1.1 Dataset:* Table 1 shows details about the properties of the dataset used for training, validation, and testing of SEER. Our dataset is built on top of Defects4J [1], which consists of 835 bugs in 17 widely-used Java projects. In addition to the data collected from Defects4J, we also augmented the dataset with higher-order mutants and automatically generated Randoop [51] tests. The former helps diversify the bugs in the dataset, and the latter generates tests that either pass or fail on the newly added bugs. Finally, we removed all the assert statements from the developer-written and automatically generated tests to avoid bias in learning from them.

Table 1: Properties of the dataset.

| Projects | # Bugs | # Mutants | #Tests #Pass, #Fail (Contribution%) | Dataset Defects4J | | | | Higher-order Mutants | | | |
|---|---|---|---|---|---|---|---|---|---|---|---|
| | | | | Developer Tests | | Randoop Tests | | Developer Tests | | Randoop Tests | |
| | | | | # Pass | # Fail | # Pass | # Fail | # Pass | # Fail | # Pass | # Fail |
| Compress | 47 | 24,322 | 30,753 (30.26%) | 1,214 | 22 | 2,965 | 2230 | 385 | 265 | 192 | 23,480 |
| Lang | 64 | 12,824 | 15,130 (14.89%) | 702 | 59 | 1,249 | 296 | 397 | 509 | 1,124 | 10,794 |
| Chart | 26 | 4,313 | 14,901 (14.66%) | 248 | 37 | 5,316 | 4987 | 154 | 109 | 229 | 3,821 |
| Math | 106 | 9,488 | 13,580 (13.36%) | 1,685 | 91 | 1,584 | 732 | 732 | 814 | 546 | 7,396 |
| Codec | 18 | 8,973 | 10,210 (10.05%) | 271 | 11 | 648 | 307 | 375 | 54 | 93 | 8,451 |
| Closure | 174 | 1,615 | 3,170 (3.12%) | 1,431 | 8 | 113 | 3 | 768 | 76 | 49 | 722 |
| JacksonDatabind | 112 | 581 | 2,818 (2.77%) | 2,133 | 27 | 77 | 0 | 410 | 94 | 42 | 35 |
| Time | 26 | 203 | 2,415 (2.38%) | 1,765 | 39 | 255 | 153 | 84 | 70 | 18 | 31 |
| Jsoup | 93 | 677 | 2,134 (2.1%) | 925 | 57 | 458 | 17 | 125 | 234 | 277 | 41 |
| Cli | 39 | 913 | 1,612 (1.59%) | 348 | 10 | 292 | 49 | 106 | 75 | 371 | 361 |
| Csv | 16 | 1,166 | 1,582 (1.56%) | 327 | 6 | 83 | 0 | 297 | 2 | 23 | 844 |
| JacksonCore | 26 | 941 | 1,420 (1.4%) | 331 | 15 | 105 | 28 | 320 | 163 | 136 | 322 |
| Gson | 18 | 531 | 1,386 (1.36%) | 678 | 15 | 93 | 69 | 295 | 92 | 144 | 0 |
| JxPath | 22 | 143 | 314 (0.31%) | 93 | 1 | 74 | 3 | 0 | 1 | 105 | 37 |
| Mockito | 38 | 0 | 100 (0.1%) | 56 | 39 | 5 | 0 | 0 | 0 | 0 | 0 |
| JacksonXml | 6 | 49 | 81 (0.08%) | 19 | 1 | 9 | 3 | 28 | 12 | 2 | 7 |
| Collections | 4 | 0 | 7 (0.01%) | 4 | 1 | 2 | 0 | 0 | 0 | 0 | 0 |
| Total | 835 | 66,739 | 101,613 (100%) | 12,230 | 439 | 13,328 | 8,877 | 4,476 | 2,570 | 3,351 | 56,342 |

As shown in Table 1, the final augmented dataset consists of 33, 385 passing pairs and 68, 228 failing pairs of $\langle t_i, m_i \rangle$, making a total of 101, 613 $\langle t_i, m_i \rangle$ pairs in the dataset. For Phase 1 training, we construct $\langle t_i, m_i+, m_i- \rangle$ tuples by merging passing and failing $\langle t_i, m_i \rangle$ pairs for common tests. That is, for a given test $t_i$, we find all the $m$ passing pairs and $n$ failing pairs of $\langle t_i, m_i \rangle$. If $m$ and $n$ are non-zero, we get total $m \times n$ tuples of $\langle t_i, m_i+, m_i- \rangle$. This provides us with 20, 759 tuples of $\langle t_i, m_i+, m_i- \rangle$ for Phase 1 training, divided into 90% training, 5% validation, and 5% testing instances. For Phase 2 training, we similarly divided the original dataset with 101, 613 instances represented by Table 1.

*5.1.2 Learning Module Configuration:* We implemented SEER's *Learning Module* using PyTorch [54] open-source library. Multiple factors can affect the neural models' learning process and final performance. For the loss function, which determines how well the algorithm approaches learning from the training data, we used Margin Ranking Loss (MRL) in Phase 1 training and Weighted Cross-Entropy Loss (WCEL) in Phase 2 training. MRL has been shown to outperform Cross-Entropy Loss in learning the embeddings and putting data instances of the same target class closer to each other than instances from other classes [30]. Since our embedding goal is similar, i.e., to put passing pairs of $\langle t_i, m_i \rangle$ closer to each other compared to failing ones, MRL was a reasonable loss function for learning the MUT and test embeddings.

For Phase 2 training, we chose Weighted Cross-Entropy Loss rather than Cross-Entropy Loss, which is commonly used in classification problems, since our dataset has more failing pairs of $\langle t_i, m_i \rangle$ compared to passing pairs. To enhance the performance, we utilize AdamW optimizer [44], which has been shown to outperform Adam optimizer [41] to update the network weights and minimize this loss function iteratively.

The other factors that affect the model's performance are hyperparameters and overfitting. We followed a guided hyperparameter tuning to find a configuration for the model that results in the best performance on the validation data. One of the most important hyperparameters is the learning rate, which controls how much to change the model in response to the estimated error each time the model updates weights. Choosing the learning rate is challenging as a value too small may result in a long training process that could get stuck, whereas a larger value may result in an unstable training process. The learning rate of SEER's *Learning Module* for Phase 1 and Phase 2 training are $1.34e^{-4}$ and $1.34e^{-6}$, respectively. The difference in learning rates is because Phase 2 learning is incremental compared to Phase 1, and a similar learning rate results in large, i.e., NaN, loss function values. Furthermore, we used 10-fold cross-validation to avoid overfitting and implemented early-stopping criteria to terminate the training. That is, we repeated the training/validation for 10 times on different training and validation sets and chose the model that achieved the best performance. To automatically terminate the learning, our patience level was 5 epochs, i.e., if the validation loss of the model did not improve in 5 consecutive epochs, we assumed that learning had reached an optimum level.

### 5.2 RQ1: Effectiveness

For this research question, we divided 101, 613 pairs of $\langle T, C, F \rangle$ instances in our dataset into 90% training, 5% validation, and 5% testing instances. To that end, we downsampled such instances for each project by 90%, and used the remaining if possible. The only exception was the Collections project, which we included its few instances only in the training set. We select accuracy, precision, recall, and F1 score as metrics to measure the effectiveness of SEER in predicting correct labels. Table 2 shows the result for this experiment under "SEER with embedding" columns. These results are obtained through a 10-fold cross-validation, i.e., downsampling repeated 10 times.

Table 2: Effectiveness and Generalization of SEER in predicting test labels. TP, FP, TN, and FN stands for True Positive, False Positive, True Negative, and False Negative, respectively.

| Subjects | SEER with embedding | |
|---|---|---|
|  | # Pass | # Fail |
|  | TP (%), FN (%) | TN (%), FP (%) |
| Compress | 92.53%, 7.47% | 98.18%, 1.82% |
| Lang | 81.07%, 18.93% | 92.6%, 7.4% |
| Chart | 94.74%, 5.26% | 88.03%, 11.97% |
| Math | 93.83%, 6.17% | 89.98%, 10.02% |
| Codec | 79.12%, 20.88% | 99.58%, 0.42% |
| Closure | 98.36%, 1.64% | 80.56%, 19.44% |
| JacksonDatabind* | 100%, 0% | 30.77%, 69.23% |
| Time* | 100%, 0% | 0%, 100% |
| Jsoup* | 98.81%, 1.19% | 0%, 100% |
| Cli | 96.23%, 3.77% | 60%, 40% |
| Csv | 94.59%, 5.41% | 98.21%, 1.79% |
| JacksonCore | 97.06%, 2.94% | 41.94%, 58.06% |
| Gson* | 100%, 0% | 66.67%, 33.33% |
| JxPath* | 100%, 0% | 25%, 75% |
| Mockito* | 100%, 0% | 0%, 100% |
| JacksonXml* | 100%, 0% | 0%, 100% |
| Collections* | N/A, N/A | N/A, N/A |
| Total | 93.63%, 6.37% | 92.77%, 7.23% |

Each row in Table 2 shows one of our subject projects and the percentage of instances they have correctly predicted for different versions of SEER. These results confirm **the original implementation of SEER illustrated in Figure 5 can effectively predict passing and failing labels for the test suite of each subject program, achieving** 93% **accuracy,** 86% **precision,** 94% **recall, and** 90% **F1 score**. Despite an overall good performance, SEER did not perform well on some projects (those marked by asterisks in Table 2. Our investigation showed that due to the low contribution of these projects to the dataset, the test data instances from them were either none, e.g., Collections project, or very few. Consequently, the effect size of classification was very large.

False Negatives are not big issues in our proposed technique, as SEER is interpretable and developers can quickly check the Attended tokens to verify the False Negative. To understand the reasons for False Positives, we compared the True Negative and False Positive instances from the following perspectives:

- *Test type.* The unit tests in our dataset are either developer-written tests or automatically generated by Randoop. Majority of $t_i$s for True Negative instances belonged to Randoop. However, for False Positives, half of the $t_i$s are developer-written tests while the other half are Randoop tests. As a result, there is no significant correlation between the False Positive instances and test type.
- *# Test Tokens.* The average number of test tokens for False Positive instances is 74 compared to 85 for True Negatives, which shows that SEER performs better when tests are longer. We believe that this is potentially because the representation of longer tests are unique compared to shorter tests, making it easier for the model to predict a correct label for them.
- *# MUT tokens.* The average number of MUT tokens for False Positive instances is 89 compared to 131 for True Negatives, which shows that SEER performs better when MUT's implementation has more tokens and statements. Similar to our argument about

# test tokens, short MUT sequences carry less semantic information, making it harder for the model to predict test results.
- *Bug type.* SEER correctly predicts the test results for 95% of the higher order mutants. This ratio for the real bugs from the Defects4J dataset is 80%. Given that we found no significant correlation between the number of buggy lines in False Positive and True Negative instances, we believe this happens because real bugs are unique. That is, while higher-order mutation injects bugs at different locations and considers different combinations of mutation operators, the operators are limited, making it easier for SEER to learn the bugs that involve those operators. We argue that this is not a limitation of SEER, but the dataset, which can be resolved by including more real bugs in the training dataset.

## 5.3 RQ2: Generalization

In the previous research question, we showed that SEER can effectively serve as an oracle on the unseen $\langle t_i, m_i \rangle$ pairs from the projects that were in the training dataset. In this research question, we go one step forward to investigate how SEER generalizes to predict test labels for $\langle t_i, m_i \rangle$ pairs, where $m_i$ belongs to a project that was not in the training data. To that end, we computed the contribution of each project (column "# Tests" in Table 1) and divided the dataset into high-contribution projects with contribution values greater than 10%, and low-contribution projects. We used all the instances of the projects in the high-contribution dataset for training and validation. Then, we tested the trained model on the projects in the low-contribution dataset. For this research question, we only consider the precision and recall values to evaluate the performance of SEER, as the low-contribution dataset is highly imbalanced (passing $\langle t_i, m_i \rangle$ pairs are 3.5× more than failing pairs).

Compared to the original precision and recall values computed in RQ1 (86% and 93%), the value of SEER's performance metrics on unseen projects are 77% (9% ↓) and 82% (9% ↓). Given that unseen projects have different statistical distributions compared to the projects used for training, i.e., different tokens and hence vocabularies, the performance drop is expected due to the Out of Distribution (OOD) problem [64]. Our further investigation of the misclassified instances confirmed that the model's performance

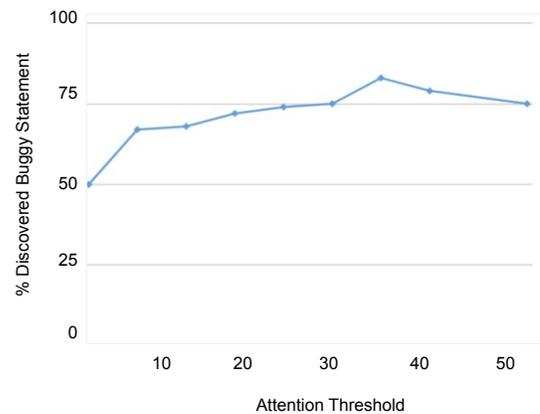

Figure 8: The percentage of attended buggy statements with respect to attention threshold

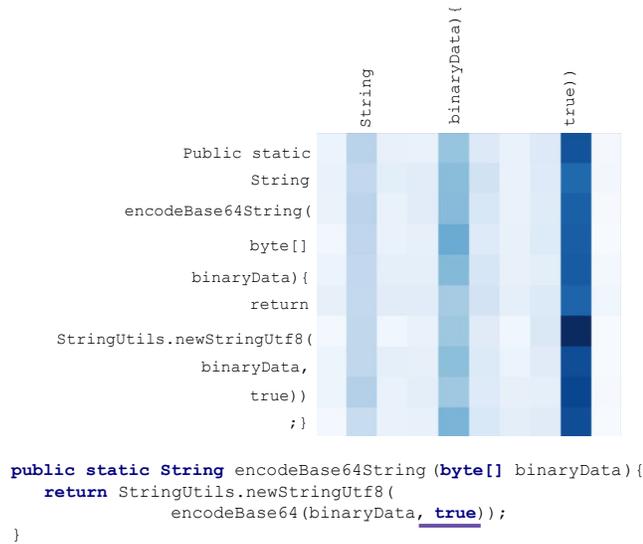

**Figure 9: Visualization of the Self Attention for the method `encodeBase64String` in Codec project, demonstrating that SEER has paid the highest attention to the buggy token, i.e., "true"**

was better on unseen projects, whose vocabularies had a higher overlap with the vocabularies of the projects used for training SEER, compared to that of for projects with less overlap in vocabularies. These results demonstrate that **SEER can achieve comparable performance on unseen projects whose vocabularies overlap with the projects for training the oracle**.

### 5.4 RQ3: Interpretation

Recall that the goal of SEER's interpretation is two-fold. First, we interpret the embedding network to verify if it has correctly learned to separate the representation of correct and buggy MUTs. More importantly, we interpret the oracle to identify which features, i.e., tokens in the MUT, and which statements were mostly attended when predicting a label. From this information, we can determine whether the attended features are relevant to the decision and whether the model's performance is valid.

*5.4.1 Attention Analysis.* Given a threshold number $k$, SEER's attention analysis algorithm (Algorithm 2) produces a set of top $k\%$ attended tokens and attended statements in the MUT. To confirm that SEER has attended to *relevant tokens* for predicting labels, we measured the percentage of the buggy statements that are among attended statements in the MUT. Specifically, we computed this metric for *true negative* test instances, i.e., $\langle t_i, m_i \rangle$ pairs for which the SEER correctly identified to fail. Figure 8 shows the percentage of buggy statements that were among the attended statements, and how this percentage changes in response to the change of threshold value. These results demonstrate that **SEER has indeed attended to relevant tokens to predict test labels, and even with a small threshold value of** 5%, **it can correctly identify** 50% **of the buggy statements in the subject MUTs**. One interesting observation here is that increasing the threshold may not result in better bug localization. For example, increasing the threshold to 40% or 50% results in non-buggy statements being among the attended statements, decreasing the contribution of buggy statements among attended statements.

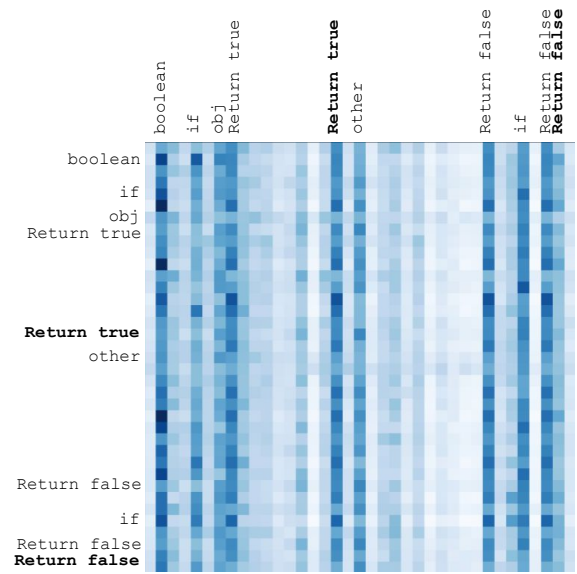

**Figure 10: Visualization of the Self Attention for the method `equals`, demonstrating that SEER has attended to buggy tokens at multiple locations**

We also manually investigated the heatmap visualization of the Self Attention matrix for the failing $\langle t_i, m_i \rangle$ pairs to qualitatively confirm if SEER has considered relevant tokens to predict labels. Figure 9 illustrates such case, where the buggy MUT is `encodeBase64String` of the Codec project. As shown in Figure 9, the bug is due to feeding an incorrect argument to `encodeBase64` method, i.e., "true" instead of "false". By looking at the heatmap visualization of the Self Attention matrix, we can see that SEER has paid the most attention to the buggy token when predicting the "fail" label for this test instance (for the sake of space and readability, we have merged some of the tokens and adjusted the weights in the heatmap visualization). As another example, where the bug is more complex and involves multiple tokens or statements, consider the buggy MUT and its corresponding heatmap [3] visualization of Self Attention in Figure 10. In this example, the return values of the two highlighted return statements are incorrect. Looking at the heatmap, we can see that the tokens of these two statements are among the most attended tokens to predict the "fail" label.

---

[3]The tokens are merged in this heatmap and only the most attended tokens are labeled.

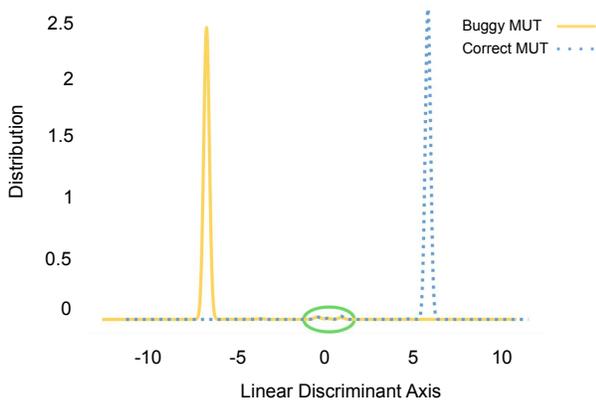

**Figure 11: Linear Discriminant Analysis (LDA) results. The yellow graph represents the distribution of buggy MUTs and the dashed blue graph represents the distribution of correct MUTs in the reduced dimension**

*5.4.2 Embedding Analysis.* Figure 11 shows the result of embedding analysis as discussed in Section 4.4.2. The blue and yellow distributions show the distinction between the embeddings of correct and buggy MUTs, respectively, after LDA dimensionality reduction. As demonstrated by this figure, the distribution of correct and buggy MUTs are almost distinct in the low-dimension space, which confirms **SEER's ability to semantically distinguish the representation of buggy and correct MUTs in the embedding space.** There are some overlapping instances near $x = 0$ between the correct and buggy MUTs distributions. Such overlap indicates that the embeddings of a few buggy and correct MUTs are close to each other in the embedding space with 200 dimensions. By manually investigating those instances, we realized that they belong to $\langle t_i, m_i \rangle$ pairs that SEER failed to predict a correct label.

## 5.5 RQ4: Performance

To answer this research question, we evaluated the time required for Phase 1 and Phase 2 training, as well as the time for testing the oracle. We ran the experiments on a Tesla T4 GPU with 16GB GDDR6 memory. For a batch size of 16 $\langle t_i, m_i \rangle$ pairs, a single epoch took 447 and 1, 745 seconds on average for training Phase 1 and Phase 2, respectively. With the patience level of five epochs for the early termination criteria, Phase 1 and Phase 2 training took 22 and 30 epochs to complete, respectively, resulting in a total of 14 hours of training. Given that the SEER is generalizable, the one-time training of the model is reasonable. After training, it takes SEER only 6.5 milliseconds on average to predict the passing or failing label for a given pair of $\langle t_i, m_i \rangle$.

## 6 RELATED WORK

State-of-the-art test oracle automation techniques can be divided into three main categories: *Implicit*, *Specified*, and *Derived* oracles. *Implicit oracle* relies on some implicit knowledge to identify whether a test passes or fails. Examples of such implicit knowledge are buffer overflow almost always yields an error, excessive CPU usage is a likely indicator of server disruptions, and unnecessary battery usage is evidence of energy defects in mobile apps. While quite effective and automated, implicit oracles can only determine the presence of limited categories of bugs.

*Specified oracle* determines the *expected* output of test execution and compares it with the actual output to decide whether a test passes or fails. To identify the expected output, these oracles require existence of formal specifications [18, 24, 28, 36, 38, 55] or contracts (pre-and post-conditions) [5, 9, 16, 17, 78] for the system under test. The performance and usability of such oracle highly depend on the availability, completeness, and quality of specifications. However, for many ever-changing software systems, specifications and contracts either do not exist or fall out of date. Even if automated techniques generate specifications, they are usually quite abstract, and inferring concrete test outputs from them is not guaranteed or is imprecise [6, 27]. State-of-the-art ML-enabled techniques alleviate such limitations by predicting meaningful specifications [48] or assert statements [72, 77, 80]. Compared to SEER, these techniques evaluate a limited set of properties related to program behavior only at a certain point the assertions [18].

*Derived oracle* decides the passing or failure of a test by distinguishing the system's correct from incorrect behavior rather than knowing the exact output. The correct and incorrect behavior can be (1) inferred from some meta-data such as *execution logs* [3, 7, 8, 22, 23, 26, 31, 42, 47, 50, 56, 63, 65, 66, 75]; (2) provided as properties of the intended functionality (metamorphic relations) [4, 11, 12, 19–21, 43, 46, 49, 57–62, 68, 69, 71, 73, 83, 85, 89, 91, 92]; or (3) checked against other versions of the software [14, 32, 67, 81, 82, 86]. Derived oracles are pragmatic, but are generally incomplete, i.e., can only identify test outputs for a subset of inputs. SEER, while considered as a derived oracle, alleviates this problem through a domain-specific embedding, i.e., semantically separating the buggy and correct code in the embedding space with respect to test results. Consequently, the neural model that serves as the oracle considers that *general* knowledge to predict a passing or failing verdicts.

## 7 CONCLUDING REMARKS

Test oracle automation has been one of the most challenging problems in the software engineering community, yet it has received less attention compared to test input generation. This paper proposed SEER, a novel DL-enabled technique to move one step forward in advancing automated test oracle constructions. SEER predicts a passing or failing verdict for a given pair of $\langle t_i, m_i \rangle$ by learning the semantic correlation between inputs and outputs from a high-quality and diverse dataset. Our experimental results show that the learned oracle is accurate and efficient in predicting test results, and generalizable to the projects it has not seen during training.

Currently, we are considering several directions for future work. Based on the promising results of our produced domain-specific representations for code and tests, we will explore its application in other software analysis tasks such as vulnerability detection, bug localization, and program repair. Also, we are planning to expand SEER to system tests. System tests are more complex and bigger than unit tests, which may entail changing the SEER's architecture to Graph Neural Networks (GNN) to better capture the code semantics and representations.